\documentclass[conference]{IEEEtran}

\usepackage{scalerel}
\usepackage[super]{nth}
\usepackage{tikz}
\usetikzlibrary{svg.path}

\definecolor{orcidlogocol}{HTML}{A6CE39}
\tikzset{
  orcidlogo/.pic={
    \fill[orcidlogocol] svg{M256,128c0,70.7-57.3,128-128,128C57.3,256,0,198.7,0,128C0,57.3,57.3,0,128,0C198.7,0,256,57.3,256,128z};
    \fill[white] svg{M86.3,186.2H70.9V79.1h15.4v48.4V186.2z}
                 svg{M108.9,79.1h41.6c39.6,0,57,28.3,57,53.6c0,27.5-21.5,53.6-56.8,53.6h-41.8V79.1z M124.3,172.4h24.5c34.9,0,42.9-26.5,42.9-39.7c0-21.5-13.7-39.7-43.7-39.7h-23.7V172.4z}
                 svg{M88.7,56.8c0,5.5-4.5,10.1-10.1,10.1c-5.6,0-10.1-4.6-10.1-10.1c0-5.6,4.5-10.1,10.1-10.1C84.2,46.7,88.7,51.3,88.7,56.8z};
  }
}

\newcommand\orcidicon[1]{\href{https://orcid.org/#1}{\mbox{\scalerel*{
\begin{tikzpicture}[yscale=-1,transform shape]
\pic{orcidlogo};
\end{tikzpicture}
}{|}}}}
\newcommand{\vect}[1]{\boldsymbol{#1}}

\usepackage{cite}
\usepackage{amsmath,amssymb,amsfonts}
\usepackage{algorithmic}
\usepackage{graphicx}
\usepackage{textcomp}
\usepackage{xcolor}
\usepackage{siunitx}
\usepackage{hyperref}
\usepackage[export]{adjustbox}
\usepackage{booktabs}
\usepackage{multirow}
\usepackage{glossaries-extra}

\setabbreviationstyle[acronym]{long-short}
\glssetcategoryattribute{acronym}{nohyperfirst}{true}

\makeatletter

\usepackage{eso-pic}
\newcommand\AtPageUpperMyright[1]{\AtPageUpperLeft{
 \put(\LenToUnit{0.1\paperwidth},\LenToUnit{-1cm}){
     \parbox{1\textwidth}{\raggedright\fontsize{9}{11}\selectfont #1}}
 }}
\newcommand{\varA}[1]{{\operatorname{#1}}}
\newcommand{\conf}[1]{
\AddToShipoutPictureBG*{
\AtPageUpperMyright{#1}
}
}

\def\BibTeX{{\rm B\kern-.05em{\sc i\kern-.025em b}\kern-.08em
    T\kern-.1667em\lower.7ex\hbox{E}\kern-.125emX}}
\begin{document}

\title{Oversampling Highly Imbalanced Indoor Positioning Data using Deep Generative Models\\}
\conf{IEEE SENSORS 2021}

\author{\IEEEauthorblockN{Fahad Alhomayani \orcidicon{0000-0002-4914-8722} and Mohammad H. Mahoor}
\IEEEauthorblockA{Department of Electrical and Computer Engineering \\
University of Denver\\
Denver, USA \\
\texttt{\href{mailto:fahad.al-homayani@du.edu}{fahad.al-homayani@du.edu}} and \texttt{\href{mailto:mmahoor@du.edu}{mmahoor@du.edu}}}
}

\newacronym{GAN}{GAN}{Generative Adversarial Network}
\newacronym{CGAN}{CGAN}{Conditional GAN}
\newacronym{ACGAN}{ACGAN}{Auxiliary Classifier Generative Adversarial Network}
\newacronym{ISTA}{ISTA}{Iterative Shrinkage and Thresholding Algorithm}
\newacronym{RSS}{RSS}{Received Signal Strength}
\newacronym{CSI}{CSI}{Channel State Information}
\newacronym{DCGAN}{DCGAN}{Deep Convolutional Generative Adversarial Network}
\newacronym{CNN}{CNN}{Convolutional Neural Network}
\newacronym{WNIC}{WNIC}{Wireless Network Interface Controller}
\newacronym{AP}{AP}{Access Point}
\newacronym{GPR}{GPR}{Gaussian Process Regression}
\newacronym{LSGAN}{LSGAN}{Least Squares Generative Adversarial Network}
\newacronym{LBS}{LBS}{Location-Based Service}
\newacronym{GNSS}{GNSS}{Global Navigation Satellite System}
\newacronym{$k$-NN}{$k$-NN}{$k$-Nearest Neighbor}
\newacronym{RP}{RP}{reference point}
\newacronym{BS}{BS}{Base Station}
\newacronym{FID}{FID}{Fréchet Inception Distance}
\newacronym{MMD}{MMD}{Maximum Mean Discrepancy}
\newacronym{IS}{IS}{Inception Score}
\newacronym{SMOTE}{SMOTE}{Synthetic Minority Oversampling TEchnique}
\newacronym{ADASYN}{ADASYN}{ADAptive SYNthetic}
\newacronym{VAE}{VAE}{Variational Autoencoder}
\newacronym{CVAE}{CVAE}{Conditional Variational Autoencoder}
\newacronym{BLE}{BLE}{Bluetooth Low Energy}
\newacronym{IoT}{IoT}{Internet of Things}
\newacronym{SVM}{SVM}{Support Vector Machine}
\newacronym{GPU}{GPU}{Graphics Processing Unit}

\maketitle

\begin{abstract}
The location fingerprinting method, which typically utilizes supervised learning, has been widely adopted as a viable solution for the indoor positioning problem. Many indoor positioning datasets are imbalanced. Models trained on imbalanced datasets may exhibit poor performance on the minority class(es). This problem, also known as the “curse of imbalanced data,” becomes more evident when class distributions are highly imbalanced. Motivated by the recent advances in deep generative modeling, this paper proposes using Variational Autoencoders and Conditional Variational Autoencoders as oversampling tools to produce class-balanced fingerprints. Experimental results based on Bluetooth Low Energy fingerprints demonstrate that the proposed method outperforms SMOTE and ADASYN in both minority class precision and overall precision. To promote reproducibility and foster new research efforts, we made all the codes associated with this work publicly available.
\end{abstract}


\section{Introduction}

Interest in indoor positioning research has substantially grown in recent years due to the multitude of applications enabled by indoor positioning, such as the \gls{IoT} \cite{macagnano2014indoor}, Indoor Location-based Services \cite{werner2014indoor}, and Ambient Assisted Living \cite{6399501}. Unlike outdoor positioning, where the \gls{GNSS} is the de facto standard for positioning, there is no universally agreed-upon solution for the indoor positioning problem. Among the techniques used for indoor positioning, location fingerprinting, or simply fingerprinting, has received the most attention because of its simplicity and ability to produce accurate positioning estimates \cite{8692423}. The concept of fingerprinting is to identify indoor spatial locations based on location-dependent measurable features (i.e., location fingerprints) collected at predefined \glspl{RP}. Examples of location fingerprints include radio frequency fingerprints (e.g., WiFi \cite{RADAR}, Bluetooth \cite{6550414}, cellular \cite{10.1007/11551201_9}), magnetic field fingerprints \cite{8626558}, and hybrid fingerprints \cite{Azizyan_2009}. Fingerprinting typically utilizes supervised learning and is inherently dependent on labeled datasets. However, often real-world indoor positioning datasets are imbalanced, meaning that the class distribution of fingerprint samples is not uniform. For example, Table \ref{imbalanced_datasets} illustrates discrepancies between the number of samples in the minority and majority classes of some publicly available indoor positioning datasets. Training on imbalanced data may result in a model biased toward the majority class(es). The techniques used to address this problem can be grouped into four main approaches: data sampling \cite{chawla2002smote}, algorithmic modification \cite{10.1145/502512.502540}, cost-sensitive learning \cite{chawla2008automatically}, and ensemble learning \cite{5978225}. This paper deals with data sampling and, in particular, with oversampling data techniques. To the best of our knowledge, no study exists that investigates the problem of imbalanced data in the context of indoor positioning. The main contribution of this paper is the application of a \gls{VAE} \cite{Kingma2014} and a conditional variant, referred to as a \gls{CVAE} \cite{sohn2015learning}, on a highly imbalanced indoor fingerprinting dataset. By using various performance evaluation metrics, the achieved results are compared to those obtained by two state-of-the-art oversampling methods known as \gls{SMOTE} \cite{chawla2002smote} and \gls{ADASYN} sampling \cite{4633969}. The remainder of this paper is organized as follows: Section \hyperref[sec2]{II} describes the dataset used in this study, Section \hyperref[sec3]{III} outlines the experimental setup, and Section \hyperref[sec4]{IV} discusses the results and future research directions.

\begin{table}[!t]
\caption{Examples of imbalanced indoor positioning datasets}
\begin{adjustbox}{width=0.8\columnwidth,center=\columnwidth}
\resizebox{\columnwidth}{!}{
\footnotesize
\centering
\begin{tabular}{@{}m{2.9cm}m{1.1cm}m{1cm}m{1cm}m{1.1cm}@{}} 
\toprule\bfseries Dataset & \bfseries Type & \bfseries Minority & \bfseries Majority & \bfseries Ratio\\  
\midrule
Dataset described in \cite{data2040032} & WiFi & \si{1}& \si{2} & \si{1:2}\\

Dataset described in \cite{data4010012} & BLE & \si{36}& \si{78} & $\approx$ \si{1:2}\\

Dataset described in \cite{data5030067} (fingerprints from \nth{1} deployment) & BLE & \si{240}& \si{1,680} & $\approx$ \si{1:7}\\

Miskolc IIS \cite{toth2016miskolc} & Hybrid & \si{18}& \si{208} & $\approx$ \si{1:12}\\

Dataset described in \cite{mohammadi2018semisupervised} & BLE & \si{2}& \si{34} & \si{1:17}\\

Dataset described in \cite{barsocchi2016multisource} & Magnetic & \si{17}& \si{404} & $\approx$ \si{1:24}\\ 
UJIIndoorLoc \cite{torres2014ujiindoorloc} & WiFi & \si{2}& \si{139} & $\approx$ \si{1:70}\\

Dataset described in \cite{data3020013} & LoRaWAN & \si{1}& \si{398} & \si{1:398}\\ 
\bottomrule
\end{tabular}
}
\end{adjustbox}
\label{imbalanced_datasets}
\end{table}

\begin{figure*}[!t]
\centering
\includegraphics[width=0.8\textwidth]{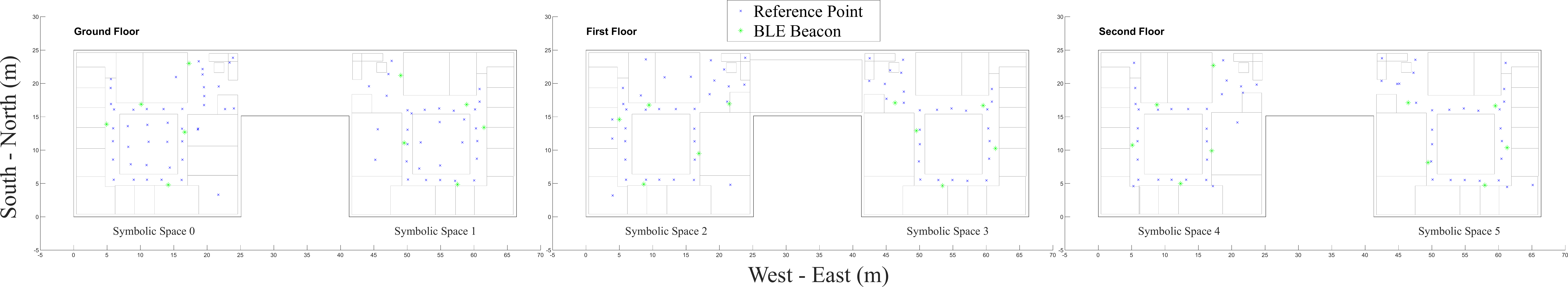}
\caption{A graphical representation of the collection environment showing \num{2}D floor plans, \glspl{RP}, and beacon locations}
\label{floorplan}
\end{figure*}

\section{Dataset Description}
\label{sec2}

\begin{figure}[!b]
\centering
\includegraphics[width=0.5\columnwidth]{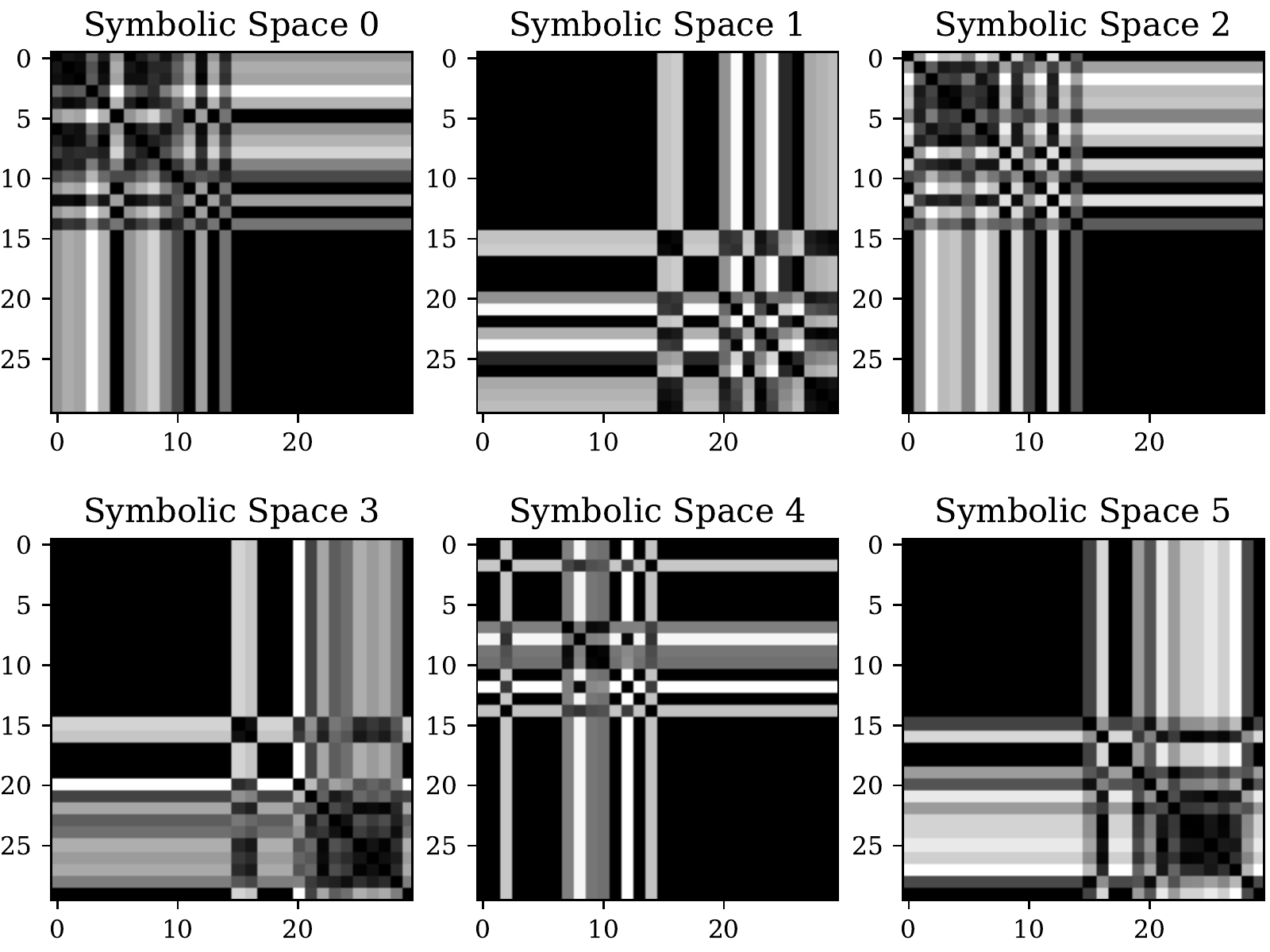}
\caption{Examples of fingerprints transformed into recurrence plots}
\label{samples}
\end{figure}

Aranda \textit{et al}. \cite{data5030067} introduced the dataset used in this study and made it publicly available. We chose this dataset because it is composed of \gls{BLE} fingerprints. \gls{BLE} is a recently introduced low-power communication protocol. It was designed with the \gls{IoT} in mind, so it has received widespread adoption in indoor positioning applications \cite{8242361}. The data we used was collected from a three-story Physics Department building. Each floor was comprised of two same-sized cubic structures joined by a hallway. Ten multi-slot \gls{BLE} beacons were deployed per floor, and three different smartphones were used to collect fingerprints at various \glspl{RP}. This paper is concerned with users' locations expressed symbolically instead of physically, also known as symbolic positioning \cite{doi:10.1080/17489725.2018.1455992}. Therefore, we treated each cubic structure on each side of a floor as an independent symbolic space. Since each symbolic space has different \gls{BLE} signal propagation characteristics, it can be considered a unique class, and the symbolic positioning problem can be cast as a classification problem. We preprocessed the dataset to exclude any samples collected outside of the cubic structures and create an initially balanced dataset. Additionally, to account for differences in beacon transmission powers resulting from multi-slot configuration, we transformed all fingerprints into recurrence plots according to \eqref{eq1}: \begin{equation}
\begin{split}
\vect{x}=[x_1,x_2,\cdots,x_n];  R_{i,j}=|x_i - x_j|; \\  \vect{x} \in  \mathbb{R}^n: \{x_i,x_j \in \mathbb{R} \mid 0 \leq x_i,x_j \leq 1\}
\end{split}
\label{eq1}
\end{equation}
where $\vect{x}$ is a fingerprint vector of dimension $n$; $x_i,x_j$ are standardized \gls{RSS} measurements corresponding to beacons $i$ and $j$, respectively; and $R_{i,j}$ represents the distance between two \gls{RSS} measurements. After preprocessing, the balanced dataset contained a total of \si{8,500} samples per symbolic space. We allocated \SI{80}{\percent} of those for training and the remaining \SI{20}{\percent} for testing. Fig. \ref{floorplan} presents a \num{2}D scheme depicting the collection environment, \glspl{RP}, and beacon locations, while Fig. \ref{samples} displays the recurrence plot of a randomly selected fingerprint from each symbolic space.

\section{Experimental Setup}
\label{sec3}

\begin{table*}[!h]
\caption{Downstream classifier results (relative to the baseline)} 
\begin{adjustbox}{width=0.55\textwidth,center=\textwidth}
\resizebox{\textwidth}{!}{
\centering
	\begin{tabular}{@{}c*{10}{S[table-format=-3.4]}@{}}
	\toprule
	& \multicolumn{3}{r}{\bfseries Minority} & \multicolumn{3}{r}{\bfseries Majority} & \multicolumn{3}{r}{\bfseries Overall} \\
	\cmidrule(lr){3-5} \cmidrule(lr){6-8} \cmidrule(lr){9-11}
	{\bfseries Minority Classes} & {\bfseries Method} & {Precision} & {Recall} & {F1-score} & {Precision} & {Recall} & {F1-score} & {Precision} & {Recall} & {F1-score} \\
	\cmidrule(lr){1-1} \cmidrule(lr){2-2} \cmidrule(lr){3-3} \cmidrule(lr){4-4} \cmidrule(lr){5-5} \cmidrule(lr){6-6} \cmidrule(lr){7-7} \cmidrule(lr){8-8} \cmidrule(lr){9-9} \cmidrule(lr){10-10} \cmidrule(lr){11-11} 
	1 & {SMOTE} & {-0.1597} & {11.0763} & {7.3103} & {0.048} & {-0.0153} & {\bfseries0.0255} & {0.0049} & {0.0813} & {0.1511}  \\ 
	& {ADASYN} & {-0.1628} & {\bfseries11.3157} & {\bfseries7.419} & {\bfseries0.0486} & {-0.0164} & {0.0254} & {0.0047} & {\bfseries0.0822} & {\bfseries0.1529}  \\
	& {VAE} & {\bfseries-0.0572} & {9.6271} & {5.9637} & {0.0297} & {-0.0537} & {-0.0444} & {\bfseries0.0117} & {0.0305} & {0.0592}  \\
	& {CVAE} & {-0.0775} & {2.6687} & {2.2359} & {0.0106} & {\bfseries0.0001} & {0.0078} & {-0.0077} & {0.0234} & {0.0462}  \\
	\cmidrule(lr){1-11}
	2 & {SMOTE} & {-0.1612} & {2.6073} & {1.7459} & {\bfseries0.0552} & {-0.0731} & {0.0137} & {-0.0295} & {\bfseries0.1778} & {\bfseries0.2649}  \\ 
	& {ADASYN} & {-0.1619} & {\bfseries2.6083} & {\bfseries1.7461} & {0.0538} & {-0.0742} & {0.0129} & {-0.0306} & {0.1769} & {0.2643}  \\
	& {VAE} & {\bfseries-0.0363} & {0.5386} & {0.5013} & {0.0128} & {\bfseries-0.0001} & {0.0123} & {\bfseries-0.0052} & {0.0504} & {0.0832}  \\
	& {CVAE} & {-0.0953} & {0.5552} & {0.4981} & {0.016} & {-0.0007} & {\bfseries0.0141} & {-0.0268} & {0.0514} & {0.0843}  \\
	\cmidrule(lr){1-11}
	3 & {SMOTE} & {-0.1863} & {3.9258} & {\bfseries2.4359} & {\bfseries0.234} & {-0.1229} & {0.101} & {-0.0323} & {\bfseries0.3697} & {\bfseries0.6369}  \\ 
	& {ADASYN} & {-0.1876} & {\bfseries3.9276} & {2.4334} & {\bfseries0.234} & {-0.0663} & {\bfseries0.1318} & {-0.0332} & {0.3692} & {0.636}  \\
	& {VAE} & {\bfseries-0.0453} & {1.533} & {1.2109} & {0.0703} & {-0.0039} & {0.0508} & {\bfseries-0.0029} & {0.1637} & {0.305}  \\
	& {CVAE} & {-0.077} & {1.3086} & {1.0386} & {0.0672} & {\bfseries-0.0029} & {0.0473} & {-0.0242} & {0.1401} & {0.2644}  \\
	\cmidrule(lr){1-11}
	4 & {SMOTE} & {-0.0907} & {\bfseries2.1388} & {\bfseries1.5017} & {\bfseries0.5263} & {-0.1097} & {\bfseries0.2738} & {0.024} & {\bfseries0.5032} & {\bfseries0.8718}  \\ 
	& {ADASYN} & {-0.0932} & {2.1385} & {1.4979} & {0.5242} & {-0.1139} & {0.2703} & {0.0216} & {0.5001} & {0.8682}  \\
	& {VAE} & {0.0282} & {0.9697} & {0.8676} & {0.1912} & {-0.0064} & {0.1279} & {0.0584} & {0.2597} & {0.4881}  \\
	& {CVAE} & {\bfseries0.0618} & {0.7553} & {0.6843} & {0.1363} & {\bfseries-0.0045} & {0.0927} & {\bfseries0.0756} & {0.2027} & {0.3808}  \\
	\cmidrule(lr){1-11}
	5 & {SMOTE} & {0.012} & {\bfseries0.2202} & {\bfseries0.2808} & {\bfseries0.1315} & {0.0601} & {\bfseries0.3246} & {0.0283} & {\bfseries0.1845} & {\bfseries0.2881}  \\ 
	& {ADASYN} & {0.0084} & {0.2119} & {0.2724} & {0.1245} & {\bfseries0.0638} & {0.3231} & {0.0242} & {0.1789} & {0.2809}  \\
	& {VAE} & {0.0705} & {0.1046} & {0.1461} & {0.0419} & {0.0477} & {0.1499} & {0.0666} & {0.0919} & {0.1468}  \\
	& {CVAE} & {\bfseries0.0782} & {0.1008} & {0.1433} & {0.0555} & {0.0423} & {0.1457} & {\bfseries0.0751} & {0.0877} & {0.1438}  \\

	\bottomrule  
\end{tabular} 
} 
\end{adjustbox}
\label{table2}
\end{table*}

\begin{figure}[!t]
\centering
\includegraphics[width=0.4\columnwidth]{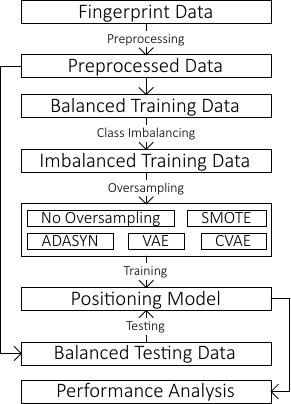}
\caption{Scheme of the experimental setup}
\label{Flowchart}
\end{figure}

Q. Li \textit{et al}. \cite{8891678} demonstrated how site surveying costs can be reduced through the incorporation of \gls{GAN}-synthesized fingerprints. In contrast, this paper addresses the problem of imbalanced fingerprint datasets using \glspl{VAE}/\glspl{CVAE}. In particular, our approach is inspired by applying deep generative models for data oversampling in domains such as fraud detection \cite{fajardo2018vos} and image processing \cite{FAJARDO2021114463}. We assessed the performance of \glspl{VAE} and \glspl{CVAE} by creating imbalanced versions of the training set. We applied these models to generate synthetic fingerprints of the minority symbolic space(s) so that all symbolic spaces are equally represented (i.e., an artificially balanced training set is created). Since we are interested in highly imbalanced data \cite{ramentol2012smote}, we set the imbalance ratio to \si{1:100} using random downsampling. We used the artificially balanced training set to train a downstream classifier that acted as a positioning model that distinguished between different symbolic spaces. For this purpose, we chose a \gls{SVM} since \glspl{SVM} are extensively used in indoor positioning \cite{doi:10.1080/17489725.2020.1817582}. We used the \texttt{scikit-learn} implementation of \gls{SVM} \cite{pedregosa2011scikit}, with default parameters that were kept fixed for all experiments. We used the testing set, which is well-balanced and remains the same for all experiments, to quantify the performance of the classifier according to metrics Precision, Recall, and F\num{1}-score as defined in \cite{sklearn_metrics}. The aim is to determine whether \glspl{VAE} and \glspl{CVAE} can learn the distribution of the minority symbolic space(s) to generate synthetic fingerprints that promote enhancements in the classifier’s performance. The performance of the classifier trained on the imbalanced version of the training set serves as the baseline. Performance results are expressed as a relative change compared to the baseline as calculated by \eqref{imb}: \begin{equation}
\varA{C_\Phi} = \frac{\varA{\Psi_\Phi}-\varA{\Psi_{IMBALANCED}}}{\varA{\Psi_{IMBALANCED}}}
\label{imb}
\end{equation}
where $\varA{C_\Phi}$ is the relative change for a performance metric $\varA{\Psi}$ obtained using an oversampling technique $\varA{\Phi}$. Since there is a total of six symbolic spaces, we performed a total of five experiments. Each experiment corresponds to a different number of minority symbolic spaces ranging from \num{1} to \num{5}. We conducted three trials for a given number of minority spaces (i.e., three imbalanced sets are constructed in which the spaces constituting a set are randomly chosen). For example, the experiment dealing with five minority spaces is composed of sets $\{0,1,2,3,5\}$, $\{0,1,3,4,5\}$, and $\{0,1,2,3,4\}$. The result is determined by averaging performance over all the trials. Table \ref{table2} presents the results of the experiments and compares them to those achieved by \gls{SMOTE} and \gls{ADASYN} as implemented in the \texttt{imbalanced-learn} library \cite{lemaitre2017imbalanced}. We used the default parameters for \gls{SMOTE} and \gls{ADASYN} and we kept them fixed for all the experiments. Similarly, \gls{VAE} and \gls{CVAE} architecture and hyperparameters implemented using \texttt{Keras} \cite{Keras} were kept fixed for all the experiments. The model specifications for \gls{VAE} and \gls{CVAE} are provided in Table \ref{table_sp} and a general scheme of the experimental setup is presented in Fig. \ref{Flowchart}. 

\section{Discussion and Conclusion}
\label{sec4}

The results in Table \ref{table2} show that, in all the experiments, using synthetic fingerprints generated by \gls{VAE}, \gls{CVAE}, \gls{SMOTE}, and \gls{ADASYN} all lead to an improved $\varA{F1-score}$ for the minority symbolic space(s) compared with classifiers trained on imbalanced datasets. Moreover, in all the experiments, every oversampling technique also resulted in a better $\varA{F1-score}$ for the majority symbolic space(s) and all spaces overall. This suggests that these oversampling techniques can enhance a classifier's overall learning ability, given that improvements are not isolated to the performance on the minority space(s). Finally, in general, \gls{SMOTE} and \gls{ADASYN} outperform \gls{VAE} and \gls{CVAE}. However, unlike \gls{VAE} and \gls{CVAE}, \gls{SMOTE} and \gls{ADASYN} are algorithms specifically designed to handle imbalanced data. Additionally, we expect that by fine-tuning \gls{VAE} and \gls{CVAE} architecture and hyperparameters, we can achieve comparable results to, if not better than, those obtained by \gls{SMOTE} and \gls{ADASYN}. Confirming this conjecture is a topic for future research. Furthermore, as part of future research, we intend to undertake a more in-depth analysis of the results to answer questions such as “Why does \gls{VAE} generally produce better overall $\varA{F1-scores}$ than \gls{CVAE}?” and “Why does \gls{VAE} yield better minority space $\varA{Precision}$ and overall $\varA{Precision}$ when the minority spaces represent \SI{50}{\percent} or less of the overall spaces, while \gls{CVAE} performs better on these metrics when the minority spaces represent over \SI{50}{\percent} of the overall spaces?”. In addition, we would like to apply \gls{VAE} and \gls{CVAE} to other fingerprint types and investigate the effectiveness of other deep generative models such as \glspl{GAN} and \glspl{CGAN} for oversampling fingerprint data. Computing scripts associated with this work are publicly available in our GitHub repository \cite{GitHublink}.

\begin{table}[!h]
\caption{\gls{VAE}/\gls{CVAE} specifications. The code for \gls{VAE} and \gls{CVAE} is inspired by \cite{keras_vae} and \cite{atienza2018advanced}, respectively, and executed on Google Colab in a \gls{GPU} runtime. } 
\begin{adjustbox}{width=\columnwidth,center=\columnwidth}
\resizebox{\columnwidth}{!}{
\begin{tabular}{|p{0.7cm} |p{4.3cm}|p{1.4cm}|p{0.7cm}|p{1.4cm}| p{0.8cm}|p{1.2cm}|}
\hline
\bfseries Order & \bfseries Layer type  & \bfseries Output size & \bfseries Filters & \bfseries Kernel size & \bfseries Strides & \bfseries Activation \\
\hline
\hline
\multicolumn{7}{|c|}{\gls{VAE} (encoder)} \\
\hline
\num{1} & Input (recurrence plot) & (\num{30},\num{30}) & - & - & - & - \\
\num{2} & Convolution & (\num{15},\num{15}) & \num{8} & (\num{4},\num{4}) & (\num{2},\num{2}) & ReLu \\
\num{3} & Convolution & (\num{8},\num{8}) & \num{16} & (\num{4},\num{4}) & (\num{2},\num{2}) & ReLu \\
\num{4} & Convolution & (\num{8},\num{8}) & \num{16} & (\num{4},\num{4}) & (\num{2},\num{2}) & ReLu \\
\num{5} & Flatten &-&-&-&-&-\\
\num{6} & Dense &\num{8}&-&-&-&ReLu\\
\num{7}(a) & Dense ($\mu$)&\num{2}&-&-&-&Linear\\
\num{7}(b) & Dense ($\sigma$)&\num{2}&-&-&-&Linear\\
\hline
\multicolumn{7}{|c|}{\gls{VAE} (decoder)} \\
\hline
\num{1} & Input (sample from distribution)& \num{2} & - & - & - & - \\
\num{2} & Dense & \si{1,024} & - & - & - & ReLu \\
\num{3} & Reshape & (\num{8},\num{8},\num{16}) & - & - & - & - \\
\num{4} & Deconvolution & (\num{15},\num{15}) & \num{16} & (\num{4},\num{4}) & (\num{2},\num{2}) & ReLu \\
\num{5} & Deconvolution & (\num{30},\num{30}) & \num{8} & (\num{4},\num{4}) & (\num{2},\num{2}) & ReLu \\
\num{6} & Deconvolution (recurrence plot)& (\num{30},\num{30}) & \num{1} & (\num{3},\num{3}) & (\num{1},\num{1}) & Sigmoid \\
\hline
\hline
\multicolumn{7}{|c|}{optimizer: Adam ($lr=\num{1e-4}$); batch size: \num{23}; objective function: binary cross-entropy $+$ Kullback–Leibler divergence} \\
\hline
\hline
\multicolumn{7}{|c|}{\gls{CVAE} (encoder)} \\
\hline
\num{1}(a) & Input (recurrence plot) & (\num{30},\num{30}) & - & - & - & - \\
\num{1}(b) & Input (label) & \num{6} & - & - & - & - \\
\num{2} & Dense & \si{900} & - & - & - & Linear \\
\num{3} & Reshape & (\num{30},\num{30},\num{1}) & - & - & - & - \\
\num{4} & Concatenate (recurrence plot \& label)& (\num{30},\num{30},\num{2}) & - & - & - & - \\
\num{5} & Convolution & (\num{15},\num{15}) & \num{16} & (\num{4},\num{4}) & (\num{2},\num{2}) & ReLu \\
\num{6} & Convolution & (\num{8},\num{8}) & \num{32} & (\num{4},\num{4}) & (\num{2},\num{2}) & ReLu \\
\num{7} & Flatten &-&-&-&-&-\\
\num{8} & Dense &\num{16}&-&-&-&ReLu\\
\num{9}(a) & Dense ($\mu$)&\num{2}&-&-&-&Linear\\
\num{9}(b) & Dense ($\sigma$)&\num{2}&-&-&-&Linear\\
\hline
\multicolumn{7}{|c|}{\gls{CVAE} (decoder)} \\
\hline
\num{1}(a) & Input (sample from distribution)& \num{2} & - & - & - & - \\
\num{1}(b) & Input (label) & \num{6} & - & - & - & - \\
\num{2} & Concatenate (sample \& label) & \num{8} & - & - & - & - \\
\num{3} & Dense &\si{2,048}&-&-&-&ReLu\\
\num{4} & Reshape & (\num{8},\num{8},\num{32}) & - & - & - & - \\
\num{5} & Deconvolution & (\num{15},\num{15}) & \num{32} & (\num{4},\num{4}) & (\num{2},\num{2}) & ReLu \\
\num{6} & Deconvolution & (\num{30},\num{30}) & \num{16} & (\num{4},\num{4}) & (\num{2},\num{2}) & ReLu \\
\num{7} & Deconvolution (recurrence plot)& (\num{30},\num{30}) & \num{1} & (\num{4},\num{4}) & (\num{1},\num{1}) & Sigmoid \\
\hline
\hline
\multicolumn{7}{|c|}{optimizer: Adam ($lr=\num{1e-4}$); batch size: \num{64}; objective function: binary cross-entropy $+$ Kullback–Leibler divergence} \\

\hline
\end{tabular}
}
\end{adjustbox}
\label{table_sp}
\end{table}

\bibliographystyle{IEEEtran}
\bibliography{Bibliography}

\end{document}